\documentstyle[psfig]{article}

\newcommand{\etal} {{\em et al.}}

\begin{document}
\title{Real-Time Spectroscopy of Gravitational
Microlensing Events - Probing the Evolution of the Galactic Bulge}
\author{D. Lennon$^1$, S. Mao$^2$, J. Reetz$^1$, T. Gehren$^1$, 
L. Yan$^3$, A. Renzini$^3$ \\
$^1$ Universit\"ats-Sternwarte M\"unchen \\
$^2$ Max-Planck Institut f\"ur Astrophysik \\
$^3$ ESO
}
\date{}
\maketitle

\section{Introduction}

Gravitational microlensing refers to the apparent brightening of a
background source by a lensing object located sufficiently close to 
the line of sight. This gravitational focusing effect does not require
the intervening object to be luminous, and hence has been suggested as a
way to detect astrophysical dark matter candidates in the Galactic halo 
\cite{Pac86}.
The challenges in detecting this effect are two-fold: Firstly, the probability
of a star in nearby galaxies (including our own) to be microlensed is
tiny, only one in a million. This means that millions of stars
have to be monitored and automatic data-processing is essential.
Secondly, one has to tell microlensing events apart
from many other intrinsic variations exhibited by stars. Fortunately,
the symmetric, achromatic and non-repeating nature of a microlensing 
event distinguishes itself. Indeed, both obstacles have been overcome and
the detection of microlensing has become a full enterprise \cite{Pac96}.
Many groups are currently monitoring the
Galactic bulge and the Large and Small Magellanic Clouds
for microlensing events\footnote{More information can be found 
at http://wwwmacho.anu.edu.au/ for the MACHO collaboration. Links 
to other collaborations can also be found there.}. At the
time of writing, more than two hundred microlensing candidates have
been discovered by the DUO, EROS, MACHO, and OGLE collaborations; Of
these, about fifteen are towards the LMC, one towards the SMC, 
while the rest 
are towards the Galactic bulge \cite{Alcock93, Alcock97a,
Udalski, Aubourg, Alard}.
An exciting observational advance is that 
most microlensing events can be
identified in real-time while they are still being lensed. This
allows detailed follow-up observations with much denser sampling,
both photometrically and spectroscopically. Two groups,
the PLANET and GMAN are conducting detailed photometric
follow-ups \cite{Alcock97c, Albrow}. 
Our group is engaged in spectroscopic observations
of selected microlensing targets (see section 2).

ESO has played a leading role in the spectroscopic studies of
microlensing events. The first spectroscopic confirmation of
microlensing was performed at ESO by Benetti, Pasquini and West
\cite{Benetti}, while the first spectral observations of a binary
lens event were carried out by Lennon {\em et al.} \cite{Lennon}. 
Such spectroscopic
studies are not only the strongest discriminator between variable
stars and genuine microlensing candidates but also of importance
for many other reasons.  The spectra obtained
allow detailed analysis of source properties, such as atmospheric
parameters, stellar radius and radial velocity. Accurate
stellar radii are essential to derive relative transverse
velocities, a quantity much needed in order to derive the lens masses.
Spectroscopic
studies also yield essential information for a small fraction of
more peculiar events.
For the exotic binary caustic events \cite{Mao}, spectroscopic studies can
resolve the stellar surface with very high accuracy
and provide new opportunities to study limb-darkening profiles,
well known only for the Sun. Spectral analysis can also provide
important clues to some puzzles in microlensing. For example,
there appears to be an over-abundance of long duration events.
Currently, it is not even known whether these lensed
sources belong to the disk or bulge populations. As these
populations are thought to be kinematically and chemically distinct,
a spectroscopic survey is needed to disentangle the disk and bulge 
contributions.

\begin{figure}
\psfig{figure=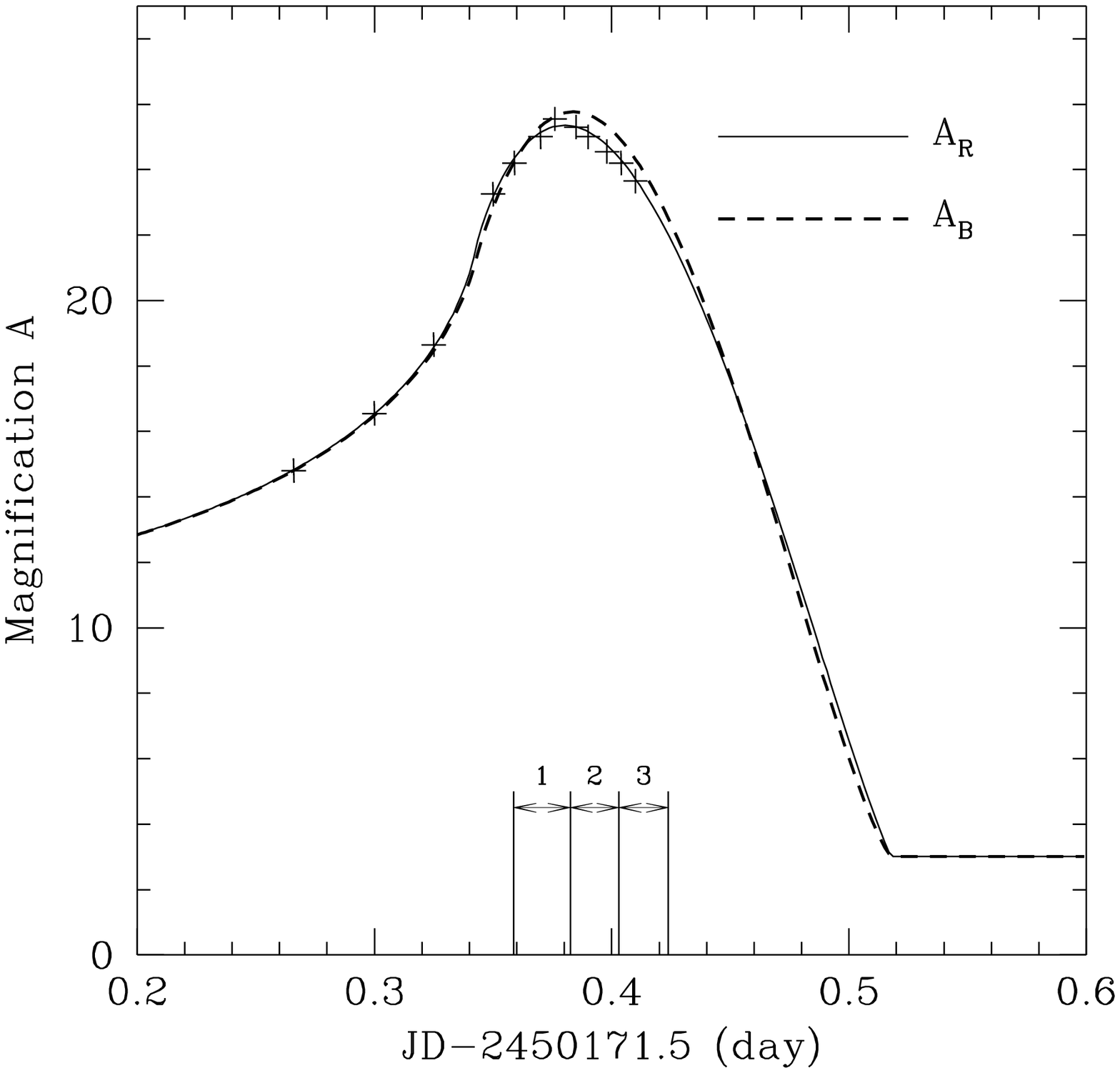,width=17.0cm}
\caption{
\label{fig:shude_fig}
Simulation of the R-band light curve for the microlensing event 96-BLG-3 near the
caustic crossing (solid line) and approximate data points (crosses).
The thick dashed line shows the prediction for the B-band.  The difference between the
magnifications in B and R is due to the variation of limb-darkening
profiles with wavelength
(here assumed to be like the Sun).
The vertical lines at bottom indicate the three 
30 minute time intervals during which our spectra were taken.
The binary nature of the lens was announced on March 28
(JD=2450171) by the MACHO collaboration \cite{Alcock96}, approximately
one day from the peak.  An earlier caustic crossing on March 25 was
also identified, as well as previous complex behaviour.
}
\end{figure}

While these aspects are of importance for understanding the observed
microlensing rate towards the Bulge, Lennon {\em et al.} \cite{Lennon}
demonstrated for the first time 
that these events presented an exciting opportunity
to investigate the formation and evolution of the Galactic Bulge itself.
The reasoning goes as follows: Arguably, the most reliable picture that we
have for the evolution of the galaxy is based upon very detailed 
abundance and kinematical studies of nearby disk
and halo cool main sequence stars, similar to the Sun.
Unfortunately, such stars in the Bulge are 
intrinsically too faint for a 3--4m class telescope to
get even a moderate resolution spectrum with good S/N.
However for the event we studied in 1996 using the NTT, the source
was a G-type dwarf undergoing a magnification by a factor of 25
at the time of observation (cf. Figure \ref{fig:shude_fig}).
The NTT was briefly the largest optical telescope
in the world!  Even with an 8--10m class telescope, such as the
VLT or Keck I/II, high resolution and high S/N spectroscopy
is out of the question for such intrinsically faint targets.
Why not make most efficient use of telescope time and carry out such
observations with the assistance of a gravitational lens?
Over several observing seasons, and with the help of
gravitational microlensing surveys, our aim is therefore to perform
a systematic spectroscopic investigation of bulge sources. We expect
that the results from this campaign will provide a fundamental insight
into the formation and evolution of the bulge of our Galaxy. In the rest
of this article we describe our first steps on this road, and summarize
the current status of the project.

\section{Program}

The feasibility of carrying out a systematic program of spectroscopic 
observations of on-going microlensing events was first discussed 
by the two lead authors early in 1996 while DJL was a visitor to
the MPIA.  
These early discussions received an unexpected boost when DJL,
while carrying out another program at the NTT telescope on La Silla,
received a telephone call from Dave Bennett of the MACHO collaboration
with the information that a binary microlensing event was {\em predicted} 
to undergo a caustic crossing during that observing run!  That event,
96-BLG-3, was duly observed by us as a target-of-opportunity
and a preliminary analysis has already been published \cite{Lennon}.
Note that for an event such as 96-BLG-3, in which the 
{\em lens} is a binary system, the light curve may differ dramatically
from the standard single lens curve, with the appearance of
spectacular spikes as the source crosses caustics or near cusps.
Extemely high amplifications may be reached during such
occurrences.
In Figure \ref{fig:shude_fig} we show schematically the timing of our
observations compared to a light curve which approximates
the behaviour of 96-BLG-3 during the relevant caustic crossing.
The MACHO team's prediction of such an exotic event 
was an impressive feat, further
strengthening our belief that on-going microlensing events
could and should be spectroscopically monitored.
Given this impetus, we therefore submitted a proposal to ESO requesting
time on the NTT for a more systematic spectroscopic investigation of
microlensing events towards the Galactic Bulge.  

We opted for the NTT for a number of important reasons.  Chief
among these was the expectation that after the `big bang', some observing
on the NTT would be offered in service mode with observations being
carried out in queue scheduled mode.  Note that
ours was not the usual kind of target-of-opportunity proposal, in the
sense that we could estimate the expected rate of discovery of
new events, as well as their probable range of magnitudes. Our
observing program could therefore be well defined except that we
would only have an advance warning of weeks or days, depending on
the event duration. Our hope was that
we could get the relevant information into the system early enough to allow
the NTT team to carry out the observations we required.  One additional
very important aspect of the NTT is that EMMI is permanently
mounted on the telescope, unlike EFOSC1 on the 3.6m telescope for example.
Our only remaining minor concern was that the relevant grism or grating
would be mounted in the instrument.  

While ESO clearly regard the NTT
as an important test of various operational and technical aspects
for the VLT, we also saw this program as a way of gaining valuable
experience (for us and for ESO) since we also hope to pursue this work
with the VLT.  We were therefore extremely gratified that the OPC
awarded us 30 hours of NTT time for this project during the
period July -- September 1997.  We were further impressed by
the professional assistance of the NTT team on La Silla
and the Data Management Division in Garching in implementing and
carrying out our program in what has been a very successful
beginning.

\section{Current Status}

At the time of writing we have data for a total of 
five events, two were observed previously as targets-of-opportunity, 
while three events
have so far been observed using the NTT and EMMI under our spectroscopic
monitoring program.  The events are described below. (We follow
the MACHO naming scheme such that 96-BLG-3 refers to event number
3 towards the Galactic Bulge in observing season 1996.)

\begin{itemize}
\item 96-BLG-3 A binary microlensing event (the lens is a binary system),
      the first to be observed spectroscopically.  It was
      observed as the source star traversed a caustic, leading to a
      very high amplification by a factor of 25 (cf. Figure
      \ref{fig:shude_fig}).
\item 97-BLG-10 Another anomalous event with evidence for caustic
      crossings, however the data reduction process is complicated 
      due to the presence of another nearby star in the aperture of the
      spectrograph. (Unlike the other events this was observed at the 
      ESO 3.6m telescope using EFOSC1 in echelle mode.)  Maximum
      amplification has been estimated as 13.3.
\item 97-BLG-26 This was a long duration high amplification event, with
      a maximum amplification of 8.0, in which 
      the source star is probably a late type sub-giant. 
\item 97-BLG-41 Another case of an anomalous event indicating
      that the lens is a multiple system.  Again this was observed
      during a caustic crossing when the source was amplified by a large
      factor.
\item 97-BLG-56 The maximimum 
      amplification for this event was also reasonably high (5.5), although
      the source is intrinsically bright and it is most likely a giant.
      The expectation here is that one may be
      able to detect finite source effect such as discussed
      in \cite{Alcock97b}.
\end{itemize}

\noindent
Due to the crowded nature of the fields used for
microlensing surveys, plus the requirement that sometimes
one is seeking to identify line-profile or continuum
slope variations, the reduction of data is a complicated
business which must be performed carefully.
This is carried out using a suite of IDL routines
developed and maintained at the Universit\"ats-Sternwarte M\"unchen.
The analysis of these data will be carried out using improved techniques
compared to those used by us in earlier work \cite{Lennon}. 

\section{The challenges}

We set ourselves the goal of deriving stellar parameters 
with typical accuracies of $\Delta$T$_{\rm eff} \le 200$\,dex,
$\Delta\log g \le 0.3$\,dex and $\Delta$[Fe/H]$ \le 0.2$.
The difficulty in achieving this objective using low resolution
spectroscopy of cool stars is illustrated 
in Figure \ref{fig:96blg3} which shows that the 
theoretical low resolution spectra ($R \approx 1300$) are only responding 
at a level of 3\% to variations in gravity and metallicity
of 0.5 and 0.3\,dex respectively.
This means that non-intrinsic features must be either eliminated or 
excluded from the fit estimation to an accuracy of better than 97\% percent.
This makes great demands on the processes of data acquisition, 
calibration, reduction and spectroscopic analysis.  In particular
we need to understand the behaviour and properties of the telescope
(NTT) and spectrograph (EMMI) used to obtain the data.  On the
analysis side we have had to develop reliable methods for the
interpretation of low resolution spectra of cool dwarfs and subgiants.
In the following we briefly discuss our techniques and some of the
problems encountered.

\subsection{Data reduction}

\begin{figure}
\psfig{figure=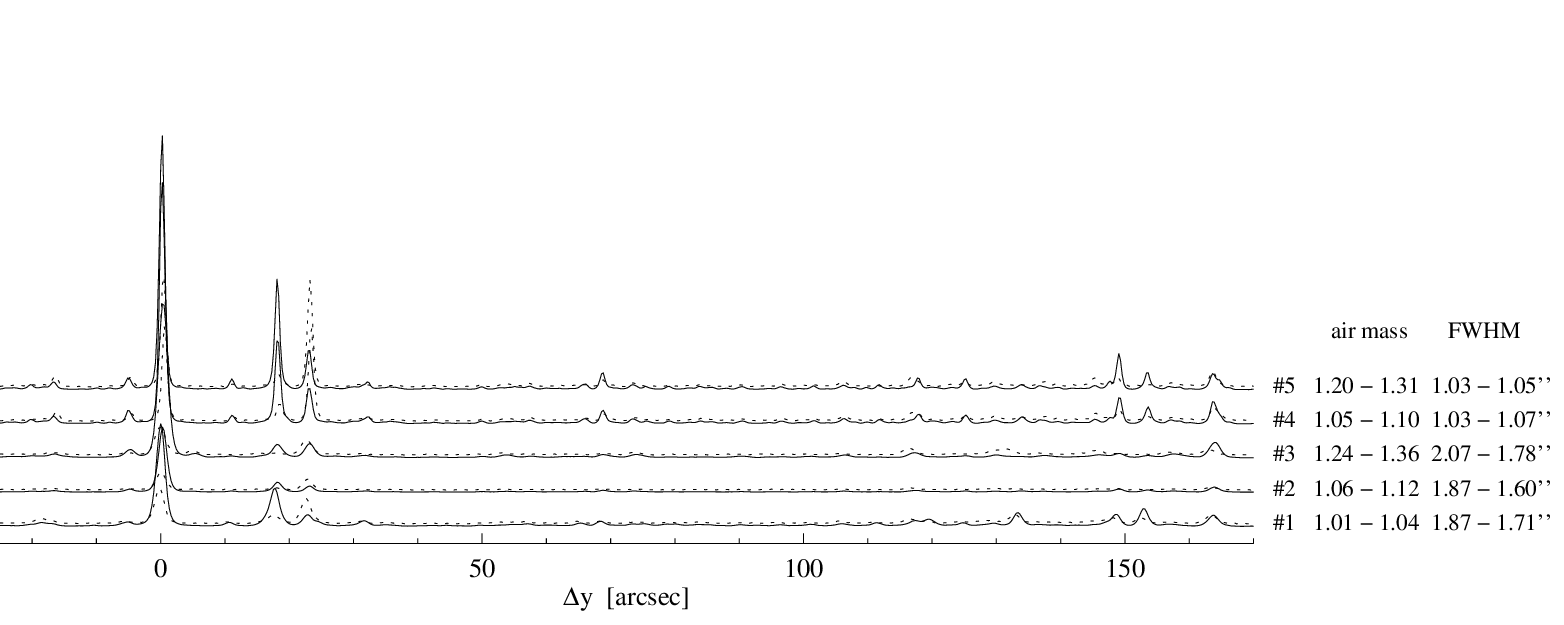,width=17.0cm}
\caption{
\label{fig:97blg56-cross}
A section of the spatial profiles of five different 97-BLG-56 exposures,
those averaged from 4380--4480\,\AA\ are dotted, 
while those averaged from 6350--6450\,\AA\ are solid lines.
The microlensed target is situated at 0 arcsec but one can see
a faint neighboring star just to the its left, well
resolved in exposures \#4 and \#5. On the right we give the air mass
and also the FWHM of the spatial profiles.
Since the seeing for exposure \#3 was best, it appears
that movement of the target along the slit may have caused additional 
broadening of the spatial profile.
}
\end{figure}

The Bulge fields are all very crowded, therefore
high {\em spatial} resolution is important to separate the target
spectrum from that of close neighbours.
Figure \ref{fig:97blg56-cross} shows spatial profiles (note that
we use a long slit) of
5 sequential exposures of 97-BLG-56; it demonstrates that in this
case a seeing FWHM smaller than one arcsec is required.
Although
photometric conditions are not required because we analyze
normalized spectra,
we need one or more additional stars on the slit to serve as
{\em differential} photometric calibrators to separate intrinsic
variations of the continuum slope from
those caused by varying transparency, seeing FWHM, airmass,
or misaligned parallactic angle.
Ideally the calibrator should sit exactly on the slit,
providing equal sensitivity to
inaccuracies in telescope pointing for both stars; of course
the angular distance between both
should be sufficiently small to minimize the differential
sensitivity to rotator inaccuracies. 
Figure \ref{fig:97blg56-cross} demonstrates that
this has not always been achieved!  Nevertheless,
we extract all spectra on the slit with a S/N$>$10, using
an optimal extraction technique. 
We are currently testing a method which allows one to
include more than a single
profile in the extraction window in order to disentangle 
spatially blended spectra.

\subsection{Spectrum Analysis}

We use
{\em line-blanked}, plane-parallel, homogeneous model atmospheres
(cf. \cite{FPFRG97}, and references 
therein).
These models (temperature and pressure structures) are
similar to those generated with ATLAS9
of R.L. Kurucz, in particular both codes use the standard
mixing-length theory to calculate the convective flux.
However we use a mixing-length parameter that is spectroscopically determined 
\cite{FPFRG97}, which has some effect on the derived effective temperature.
Line opacities are taken into account using opacity distribution functions 
from Kurucz 1992 \cite{KUR92}, scaled to account for
the fact that Kurucz' adopted solar iron abundance was overestimated.
Atomic and molecular line data originates from Kurucz, except
that most of the {\em f}-values and broadening
parameters have been adjusted such that the line profiles
fit the high resolution solar flux
atlas (\cite{KPNO84}) in the spectral regions around
${\rm H_\alpha}$, ${\rm H_\beta}$ and the Mg\,b lines 
(Figure \ref{fig:96blg3}).

We determine the {\em best fit} parameters T$_{\rm eff}$, $\log\,g$, 
[Fe/H], as well as the width of the instrumental 
profile, presently assumed to be a Gaussian.
Synthetic spectra corresponding to randomly selected sets
of parameters are interpolated from a pre-calculated grid.
A merit function for each set of parameters is 
then derived which is basically estimated as a $\chi^2$-function, 
with additional goodness-of-fit criteria used to control the weights assigned
to various pixels.
Our {\em Monte Carlo} calculation contains typically a few hundred
fit evaluations. The merits are sorted starting with the 
lowest value which corresponds to the model parameters of the {\em best}
fit, while the fit merits may be used to estimate the uncertainty.
As an example, the quality of the fit to the spectrum of 96-BLG-3 is shown 
in Figure \ref{fig:96blg3}.  
The improvement of the merit function is an important matter of 
concern in the nearest future, we will also define a physically 
based strategy to determine error boundaries.  A further refinement
currently being tested is that of deriving [Mg/Fe] and [C/Fe]
abundance ratios.

Finally, we also need to estimate the effect of the lens itself
on the perceived flux spectrum of the source since the normal
limb darkening law is to some degree distorted by the amplification
(cf. the B and R band light curves in Figure 1).
(For the present we ignore the possibility that the lens itself
contributes significantly to the observed flux.) 
For example, we know that for the sun the H$_\alpha$ line profile, which
is our primary effective temperature diagnostic,
is significantly different in center and limb spectrograms.
For 96-BLG-3 we have already computed the effect of the
lens on H$_\alpha$ and confirmed
that for this object, at the time of observation, 
the perturbation of the profile is small compared to the uncertainties
in the analysis.  However, this is of course something 
which must be considered in general, and is particularly relevant
when the source is a giant.

\subsection{Preliminary results}

\begin{figure}
\psfig{figure=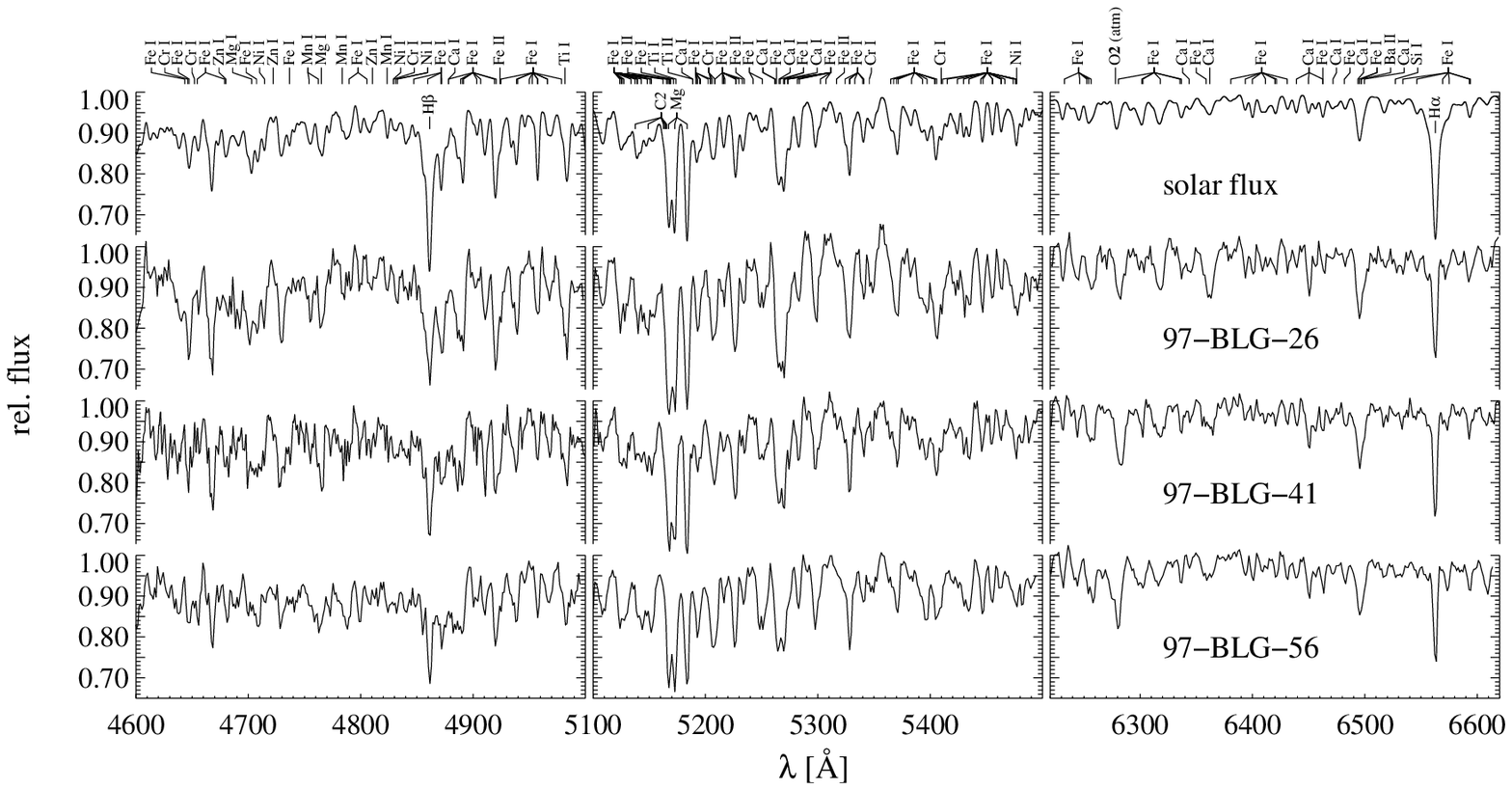,width=17.0cm}
\caption{This montage compares
the convolved solar flux spectrum (120\,km/s Gaussian) with
spectrograms of three recently observed microlensing events 
obtained with the NTT.
For all of the observations we used EMMI in RILD mode with grism\# 5 giving a 
nominal spectral resolution of 1100 for a 1 arcsec slit and 
wavelength coverage of 3985--6665\,\AA. The actual seeing limited 
resolution is 15\% higher.
The signal to noise ratio of the normalized spectra is 
$\sim 90$ (97-BLG-26), $\sim 85$ (97-BLG-41) and $\sim 200$ (97-BLG-56).
Strong spectral lines are denoted. Telluric absorption lines particularly
${\rm O_2}$ and ${\rm H_2O}$ at ${\rm H_\alpha}$ have not been eliminated, though
monitored with white dwarf exposures.\label{fig:blg-spectra}}
\end{figure}

\begin{figure}
\psfig{figure=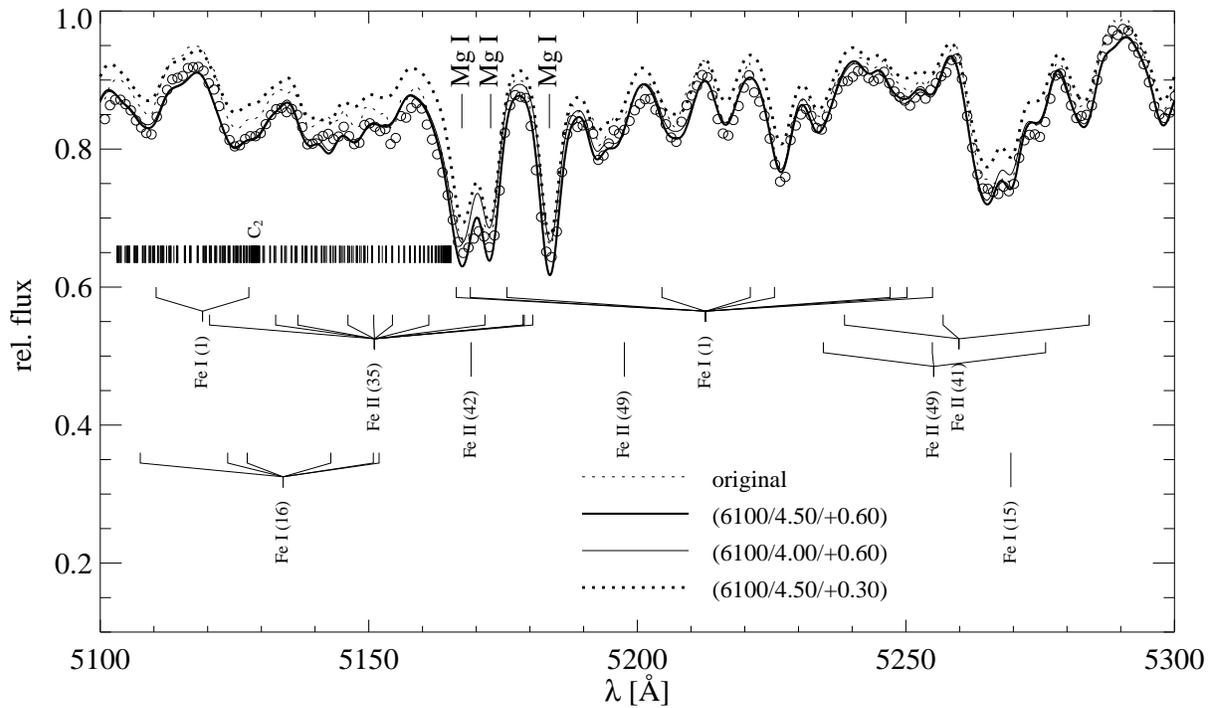,width=17.0cm}
\caption{
\label{fig:96blg3}
Comparison of synthetic spectra with the previously observed
spectrum of 96-BLG-3 (open circles). One 
improvement over our original work (thin dotted line) \cite{Lennon} 
is the inclusion of C$_2$ opacity resulting in a much improved
fit bluewards of the Mg\,{\sc i} 5167.3\,\AA\ component.
One can see that the
model ($T_{\rm eff}/\log g/[\rm{M/H}]=6100/4.50/+0.60$)
now fits all the data points quite well in this spectral range.
Strong iron multiplets are also indicated.
}
\end{figure}

In Figure \ref{fig:blg-spectra} we show a montage of spectra for the 3 targets
observed so far under the auspices of our NTT 
target-of-opportunity program.
We have derived preliminary stellar parameters for only two of these targets,
which are subgiants, since we do not yet have atmospheric models 
suitable for the analysis of giants.
The Balmer line wings presented in Figure \ref{fig:blg-spectra} indicate
effective temperatures of $\sim 5200\pm 200$\,K 
for 97-BLG-26 and $\sim 5000\pm 200$\,K for 97-BLG-41. 
Gravities are $\log\,g \sim 3.9 \pm 0.3$ and $\sim$ 3.2$\pm 0.3$, 
respectively whereas 97-BLG-26 appears to be metal-rich 
([Fe/H] $\sim 0.3$\,dex)
and 97-BLG-41 to be slightly metal-deficient ([Fe/H] $\sim -0.2$\,dex). 
Due to the high S/N we expect that the uncertainty of the 
derived metallicities does not exceed 0.3\,dex. 
We have also obtained preliminary radial velocities for
other objects falling on the long slit (cf. Figure
\ref{fig:97blg56-cross}).
The accuracy of the stellar parameters, and therefore
the derived radii and stellar masses, 
are expected to be significantly improved in a more complete analysis.

\section{Future Plans}

We plan to continue spectroscopic surveys of microlensing events,
it is clear that with a sizeable spectroscopic
sample one can learn much about the  formation and evolution of the
Galactic bulge through dynamical and stellar atmosphere studies.
The up-to-date status of this project is available at 
http://www.mpa-garching.mpg.de/$^\sim$smao/survey.html.
It will be very exciting to use the larger telescopes such as
the VLT in this work,
in fact the KECK~I has already been used to observe the
finite source size event 95-BLG-30 \cite{Alcock97b},
and indeed a more systematic spectroscopic survey
was carried out this year (Minniti, private communication).
The implications of moving to a larger aperture are obvious;
it will be possible to
resolve peculiar events with much better time (and therefore spatial)
resolution.  
More subtle events such as blending of light by the
lens may become observable with VLT due to the difference between
the lens and source radial velocities.
A series of center to limb spectra could
then be used to constrain stellar models of the poorly understood atmospheres
of cool giants/supergiants.  Microlensing candidates in the LMC and SMC, which
are typically fainter by about three magnitudes, come within reach
of spectroscopy. In particular, using the VLT and UVES,
high S/N and high resolution spectra will
be quite easily obtainable for the more strongly amplified
sources such as are discussed above. This will permit the analysis
of Bulge dwarfs with an accuracy comparable to that of their
nearby field and halo counterparts, and allow us to investigate
their chemical compositions in detail, looking at relative
abundances of light, $\alpha$-process, Fe-group, $r$-process 
and $s$-process elements.  
The VLT, given its expected
flexibility when it comes to scheduling and the wide range
of optical and IR instrumentation {\em permanently} available 
at its many foci, is ideally
suited to this kind of multi-faceted project.  

\section{Acknowledgments}

Carrying out service mode observing for a project
such as ours is not easy, but we at least want to make
sure that it is not a thankless task!  Special thanks are
due to the staff at La Silla including Gauthier Mathys,
Chris Lidman, G. van de Steene and the rest of the NTT
team.
We also thank ESO staff at Garching, particularly
Dave Silva and Albert Zijlstra in the Data Management Division
for their professional assistance. 
Of course, this work would not be possible without
the real-time alert system of the MACHO collaboration and the
photometric follow-up networks of the PLANET and GMAN
collaborations. We have also benefited from discussions with 
the EROS and OGLE collaborations.
We are indebted to Dave Bennett, Chris Stubbs and Charles Alcock
(MACHO), Penny Sackett (PLANET),
and Jim Rich (EROS) for discussions and encouragement.

\end{document}